\documentclass[preprint,a4paper,pra,showpacs]{revtex4-1}
\usepackage{amssymb}
\usepackage{amsfonts,amsbsy}
\usepackage{graphicx}

\def\e{\mathrm{e}}
\def\i{\mathrm{i}}
\def\d{\mathrm{d}}

\def\beq{\begin{equation}}
\def\eeq{\end{equation}}
\def\Re{\mathrm{Re}\,}

\def\bv{\boldsymbol{v}}

\def\E{\mathsf{E}\,}
\def\var{\mathsf{var}\,}
\def\P{\mathsf{P}}
\def\Zbf{Z^\mathrm{bf}}
\def\n{\hat e}
\def\L{\mathcal{L}}

\newcommand{\eqn}[1]{(\ref{eqn:#1})}
\newcommand{\lab}[1]{\label{eqn:#1}}
\newcommand{\inter}[1]{\quad \textrm{#1} \quad}

\newcommand{\su}[1]{^{\left(#1\right)}}

\newtheorem{proposition}{Proposition}

\def\XXint#1#2#3{{\setbox0=\hbox{$#1{#2#3}{\int}$}
\vcenter{\hbox{$#2#3$}}\kern-.5\wd0}}

\begin{document}

\title{Estimating generalised Lyapunov exponents  for products of random matrices}
\author{J. Vanneste}
\affiliation{School of Mathematics and Maxwell Institute for Mathematical Sciences \\ University of Edinburgh, Edinburgh EH9 3JZ, UK}
\email{J.Vanneste@ed.ac.uk}

\begin{abstract}
We discuss several techniques for the evaluation of the generalised Lyapunov exponents which characterise the growth of products of random matrices in the large-deviation regime. A Monte Carlo algorithm that performs importance sampling using a simple random resampling step is proposed as a general-purpose numerical method which is both efficient and easy to implement. Alternative techniques complementing this method are presented. These include the computation of the generalised Lyapunov exponents by solving numerically  an eigenvalue problem, and some asymptotic results corresponding to high-order moments of the matrix products.
Taken together, the techniques discussed in this paper provide a suite of methods which should prove useful for the evaluation of the generalised Lyapunov exponents in a broad range of applications. Their usefulness is demonstrated on particular products of random matrices arising in the study of scalar mixing by complex fluid flows. 
\end{abstract}
 
\pacs{05.45.-a,05.45.-a,47.51.+a,47.52.+j,02.70.Uu}

\maketitle

\section{Introduction}

Products of random matrices arise in many physical models, of disordered media, of wave localisation, and of chaotic dynamics in particular. The main quantity of interest is the largest Lyapunov exponent, which gives the rate of exponential growth of the products as the number $N$ of factors increases to infinity. The free energy of random Ising models, for instance, is given by the largest Lyapunov exponent of a product of matrices, as is the localisation length of some random Schr\"odinger operators. We refer the reader to the book by Crisanti, Paladin and Vulpiani \cite{cris-et-al93} for a discussion of these and other applications. 

Often, it is necessary to go beyond the almost-sure, infinite-$N$ growth of the matrix product captured by the largest Lyapunov exponent, and examine finite-$N$ fluctuations. These are characterised  by the distribution of the so-called finite-time (or finite-$N$) Lyapunov exponents, or equivalently by the generalised Lyapunov exponents $\ell(q)$, which give the growth rate of the $q$th moment of the norm of the matrix product \cite[e.g.][]{cris-et-al88,cris-et-al90,cris-et-al93,ott02}. At a mathematical level, the generalisation 
involved entails  the passage from the (mutliplicative, non-commutative) law of large numbers \cite{furs-kest,furs63,osel68} to the corresponding theory of large deviations \cite[][and references therein]{boug-lacr}.

One area of applications in which multiplicative large deviations and generalised Lyapunov exponents played a central role is the transport, mixing and reaction of constituents in complex fluid flows. In the last ten years or so, a number of results have related the macroscopic dynamics of scalars and fields in fluid flows to the large-deviation statistics of the stretching by these flows \citep[see][for an early review]{falk-et-al}. Specifically, the generalised Lyapunov exponents associated with the stretching have been found to control the decay rate of purely advected passive scalars \citep{anto-et-al,balk-foux,falk-et-al,fere-hayn,tsan-et-al05a,hayn-v05}, the spatial distribution of reacting scalars \citep{neuf-et-al,tzel-hayn10}, the reaction rate of fast reactions \citep{tsan09}, the distribution  of vorticity in certain turbulent flows \citep{reyl-et-al,tsan-et-al05b}, the clustering of inertial particles \citep{balk-et-al}, the magnetic field in kinematic dynamo models \citep{cher-et-al99}, etc. In most of these applications, the complex fluid flows are modelled by random processes which either are white in time (Kraichnan--Kazantsev flows), or consist of sequences of independent  identically distributed (iid) processes (variously termed renewing, renovating, or  innovating flows). In the latter case, the stretching is controlled by products of iid random matrices of the type considered in this paper.

In many of these applications, it is necessary to evaluate the generalised Lyapunov exponents for specific random matrices. Very few exact results are available, however. As is also the case for the usual Lyapunov exponent, given in fact by $\ell'(0)$, these are essentially limited to matrices satisfying an isotropy property that reduces the problem to scalar multiplication \citep{cohe-newm,newm86,cris-et-al93}. Thus approximations to $\ell(q)$, either perturbative or numerical, need to be obtained. Crisanti et al.\ \cite{cris-et-al93} review several techniques including Cook and Derrida's  asymptotic results for large sparse matrices \cite{cook-derr90}, the weak-disorder expansion for near-identity matrices, the replica trick (applicable when $q$ is even and positive), and the (heuristic) microcanonical estimate. Cycle expansions \citep{main92,bai08} provide yet another technique. However, these techniques are limited to special ensembles of matrices: the microcanonical and cycle-expansion estimates, for instance, are applicable to ensembles drawn from a small number of matrices. There is, therefore, a genuine need for numerical techniques that enable the estimation of $\ell(q)$ for a  broad range of matrix ensembles.  
The main aim of the present paper is to develop one such numerical technique and to demonstrate its usefulness by applying it to a few examples. 

Several of the papers on fluid mixing cited above contain numerical evaluations of the generalised Lyapunov exponents corresponding to simple renewing flows, and in particular to the alternating sine map \citep{pier94} that has become a standard tool of the field. Most of these  estimates are obtained using a straightforward Monte Carlo sampling of either the probability distribution of the finite-time Lyapunov exponents, or of the $q$th moments of the norm of the matrix product. This approach, which we refer to as brute-force Monte Carlo in what follows, is highly inefficient unless $|q|$ is small. This is because it attempts to sample events that have an exponentially small probability as $N \to \infty$. Clearly, what is needed is some form of importance sampling, which focuses the computational effort on the realisations dominating the estimate of $\ell(q)$.  
We propose and test a simple algorithms that has this property. This algorithm, which we call Resampled Monte Carlo (RMC), falls in the category of sequential importance-sampling \citep{liu01} or `go-with-the-winners' strategies \citep{gras02} used extensively in statistical physics and elsewhere; it consists of a simple modification of the brute-force computation adding a (random) resampling step which drastically reduces the sample variance. As a result, it yields accurate estimates of $\ell(q)$ with ensembles that are orders of magnitude smaller than those required for the brute-force estimation. The algorithm is very close to the cloning/pruning algorithm recently developed to estimate large-deviation statistics of more general Markov chains \citep{giar-et-al,leco-tail07} and of Lyapunov exponents in Langevin dynamics \citep{tail07,tail-kurc07}. However, our focus on products of matrices leads to an algorithm that is particularly simple to implement and to analyse. 

Recently, Haynes and Vanneste \cite{hayn-v05} used an alternative approach to the brute-force Monte Carlo sampling for the evaluation of $\ell(q)$ for the alternating sine flow \citep[see also][]{tzel-hayn10}. This approach relates $\ell(q)$ to the eigenvalue of an (infinite-dimensional) eigenvalue problem that can be discretised and solved numerically, at least for $2 \times 2$ and perhaps $3 \times 3$ matrices. We review this approach, first to compare its results with those of our RMC algorithm, but mostly because the eigenvalue problem can be used to derive interesting properties of $\ell(q)$. One such property relates the function $\ell(q)$ associated with an ensemble of matrices $A$ to the corresponding function associated with the complementary ensemble of matrices $A^{-1}/|\det A|^{1/q}$.  This relationship is of great practical interest since considering $A^{-1}/|\det A|^{1/q}$ instead of $A$ can lead to more accurate estimates of $\ell(q)$ for some value of $q$. We demonstrate the usefulness of this observation in some examples.  

For large $|q|$, $\ell(q)$ is controlled by exceedingly rare realisations of the matrix products, and hence it is difficult to estimate reliably using Monte Carlo numerical methods, even with importance sampling. An alternative, which we pursue in this paper, is to take advantage of the largeness of $|q|$ to derive asymptotic estimates. Starting with the eigenvalue problem and using a WKB ansatz, we obtain the asymptotics of $\ell(q)$ for ensembles of bounded matrices and for matrices with (not necessarily independent) Gaussian entries. These asymptotic estimates, together with the RMC method, the eigenvalue formulation, and the replica approach (which we briefly discuss) provide a suite of methods which should prove useful for the evaluation of the generalised Lyapunov exponents of products of random matrices arising in a broad range of applications.

The plan of the paper is as follows. In section \ref{gle}, we review the definition of the generalised Lyapunov exponents $\ell(q)$ and their connection with the large-deviation distribution of the finite-time Lyapunov exponent. We also derive the eigenvalue problem from which $\ell(q)$ can be inferred, and we use it to relate $\ell(q)$ obtained for the matrices $A$ to its counterpart obtained for the matrices $A^{-1}/|\det A|^{1/q}$. The RMC algorithm is presented and analysed in section \ref{rmc}; there we show that the algorithm leads to an unbiased estimate for the $q$th moment of the matrix product, and we derive an expression for the variance of this estimate. Section \ref{other} is devoted to alternative methods for the evaluation of $\ell(q)$, namely the numerical solution of the eigenvalue problem, the replica method, and the large-$|q|$ asymptotic results. All the methods dicussed in the paper are tested on three examples of random-matrix ensembles in section \ref{ex}. The paper concludes with a discussion in section \ref{conc}. A pseudocode implementing the RMC method, and some technical derivations are relegated to three appendices.

\section{Generalised Lyapunov exponents} \label{gle}

\subsection{Definitions and basic properties} \label{sec:def}

We consider $N$ successive products of a vector $X_0 \in \mathbb{R}^{d}$ by iid random matrices $A_n \in \mathbb{R}^{d \times d}$, $n=1,2,\cdots,N$. In other words, we consider the recurrence
\beq \lab{recurrence}
X_n = A_n X_{n-1}, \quad n=1,2,\cdots,N.
\eeq
We assume that $X_0$ is determistic and normalised: $X_0=x_0$ with $\|x_0\|=1$. 
The randomness of the matrices $A_n$ implies the choice of a probability measure on $\mathbb{R}^{d \times d}$. We will not be specific as to the properties of this measure; what we have in mind, as illustrated by the examples of section \ref{ex}, are random matrices defined by a number of random parameters taken from smooth distributions such as the normal or uniform distributions.

Our focus is on the large-$N$ behaviour of $\|X_N\|$. This can be characterized by considering the generalised Lyapunov exponents
\beq \lab{glyapdef}
\ell(q) = \lim_{N \to \infty} \frac{1}{N} \log \E \| X_N \|^q,
\eeq
where $\E$ denotes the expectation over the random matrices.
Note that these exponents are independent of $x_0$ for almost all $X_0$ and realisations of the matrices $A_n$ \citep[e.g.][]{cris-et-al93,ott02}. Correspondly, the large-$N$ asymptotics of the moments of $\| X_N \|$ is given by
\beq \lab{asymEXN}
\E \| X_N \|^q \sim c_q \e^{N \ell(q)}
\eeq
for some $c_q$. Note that in the commutative case $d=1$, \eqn{asymEXN} is exact with $c_q=1$.
An alternative to the definition \eqn{glyapdef} of $\ell(q)$ that makes the independence on $x_0$ obvious is
\beq \lab{glyapdef2}
\ell(q) = \lim_{N \to \infty} \frac{1}{N} \log \E \| A_N \cdots A_1 \|^q,
\eeq
where the matrix norm is the 2-norm, so that $\| A_N \cdots A_1 \|$ is the largest singular value of $A_N \cdots A_1$.

The generalised Lyapunov function $\ell(q)$, sometimes termed free energy, obviously satisfies $\ell(0)=0$ and can be shown to be convex. It
is directly related to the statistics of $\|X_N\|$ for $N \gg 1$  \citep[e.g.][]{cris-et-al93,ott02}. These are usually described in terms of the (largest) finite-$N$ Lyapunov exponents
\beq \lab{ftlyapdef}
H_N = \frac{1}{N} \log  \| A_N \cdots A_1 \|.
\eeq
The large-deviation theory asserts that the pdf $p_N$ of $H_N$ 
is approximately 
\beq \lab{largedeviation}
p_N(h) \asymp \e^{-N g(h)},
\eeq
where $\asymp$ denotes rough asymptotic equivalence, that is,  asymptotic equivalence of the logarithms as $N \to \infty$. 
The function $g$, variously termed rate function, Cram\'er function or entropy, is convex. It attains a minimum at the Lyapunov exponent $\bar h$, which satisfies
\beq \lab{lyap}
\bar h = \lim_{N \to \infty} H_N
\eeq
for almost all realisations of the random matrices, and it can be taken such that $g(\bar h)=g'(\bar h) = 0$. 
%
Note that $g$ is in fact independent of the norm chosen for $A_N \cdots A_1$, and that the same $g$ would be obtained if \eqn{ftlyapdef} was replaced by $H_N = N^{-1} \log \| X_N \|$. Using the latter point,  Laplace's method can be applied to write
\[
\E \| X_N \|^q \asymp \int \e^{N q h} \e^{-N g(h)} \, \d h \asymp \e^{N \sup_h (q h - g(h))},
\]
and conclude from \eqn{glyapdef} that $\ell$ and $g$ are Legendre transforms of one another,
\beq \lab{legendre}
\ell(q) = \sup_h \left( q h - g(h) \right).
\eeq
(Rigorous conditions on the probability measure for the $A_n$ that guarantee that \eqn{largedeviation} and \eqn{legendre} hold are given in Ref.\ \cite{boug-lacr}.)
Since $g'(\bar h) = 0$, the Legendre relationship $\ell'(q)=h$ gives
\beq \lab{lyapderi}
\bar h = \ell'(0). 
\eeq

\subsection{Eigenvalue problem} \label{eig}

The generalised Lyapunov exponents $\ell(q)$ can be found by solving a family of eigenvalue problems parameterised by $q$. To see this, we consider
\beq \lab{un}
u_n(x) = \E f(A_n \cdots A_1 x),
\eeq
for a given function $f: \mathbb{R}^d \to \mathbb{R}$. 
We now derive a backward equation for $u_n$ by noting that
\[
u_{n+1}(x) = \E f(A_{n+1} \cdots A_1 x) = \E f(A_n \cdots A_1 A x) = \E u_n(Ax),
\]
where the last expectation involves the single matrix $A$ only. Thus, for an arbitrary  $f$, the $u_n$ satisfy the recurrence relation
\beq \lab{unrec}
u_{n+1}(x) = \E u_n(Ax), \inter{with} u_0(x)=f(x).
\eeq
In the particular case where $f(x)=\| x \|^q$, so that $u_n(x_0)=\E \| X_n \|^q $, \eqn{unrec} admits solutions of the form
\beq \lab{aa}
u_n(x) = \lambda^n \| x \|^q v(\n),
\eeq
where $\n=x/\|x\| \in S^{d-1}$ is a unit vector. The scalar $\lambda$ and function $v$ are determined by introducing \eqn{aa} into \eqn{unrec} to obtain
\beq \lab{eig}
\left(\L_q v\right)(\n)=\lambda v(\n),
\eeq
where we have introduced the linear operator $\L_q$ defined by
\beq \lab{Lq}
\left(\L_q v \right)(\n) =  \E \|A \n\|^q v \left( A \n / \| A \n \| \right).
\eeq
Comparing $\E \| X_n \|^q =u_n(x_0)$ with \eqn{glyapdef} gives the following:
\begin{proposition}
The generalised Lyapunov exponent $\ell(q)$ is the logarithm of the largest eigenvalue $\lambda_1$ of \eqn{eig}:
\beq \lab{elllambda}
\ell(q)=\log \lambda_1. 
\eeq
\end{proposition}
Here we assume that the point of the spectrum with the largest modulus is an eigenvalue, $\lambda_1$. This can be guaranteed under certain assumptions. (See Ref.\ \cite{boug-lacr} where the eigenvalue problem \eqn{eig} is studied in order to establish central-limit and large-deviation results.)  Note that since $\L_q$ maps positive functions to positive function, $\lambda_1 > 0$. 

The characterisation \eqn{elllambda} of the generalised Lyapunov exponents is useful for a number of purposes. First, it gives a deterministic method for finding $\ell(q)$ by solving an eigenvalue problem, analytically in simple cases and numerically in less simple cases. Second, the eigenvalue formulation can be used to examine the convergence of $\log \E \| X_N \|^q$ as $N \to \infty$ and conclude, for instance, that the convergence is typically exponential, with an error proportional to $|\lambda_2/\lambda_1|^N$, where $\lambda_2$ is the second largest eigenvalue of \eqn{eig}. Third, the eigenvalue formulation makes it possible to establish some useful  properties of $\ell(q)$ which we now discuss.

In Appendix \ref{app:prop}, we show that the adjoint of $\L_q$ is the operator $\tilde \L_{-q-d}$,where $\tilde \L_q$ is defined as $\L_q$ in \eqn{eig}, but with the matrix $A$ replaced by $A^{-1}/|\det A|^{1/q}$.
We then have the following useful relationships between generalised Lyapunov exponents of the matrices $A$, $A^{-1}/|\det A|^{1/q}$ and $A^{-1}$.
\begin{proposition} \label{prop:elltm}
Let
\beq \lab{tilell}
\tilde \ell(q)=\lim_{N\to \infty} \frac{1}{N} \log  \E \frac{\|A_N^{-1} \cdots A_1^{-1}\|^q}{|\det(A_N \cdots A_1)|} \inter{and} \ell^-(q)=\lim_{N\to \infty} \frac{1}{N} \log \E {\|A_N^{-1} \cdots A_1^{-1}\|^q}.
\eeq
Then,
\begin{enumerate}
\item $\ell(q)=\tilde \ell(-q-d)$,
\item $\ell(q)=\ell^-(-q-d)$ if the matrices $A_n$ satisfy $\det A_n =1$,
\item $\ell(q)=\ell(-q-d)$ if the matrices $A_n$ are symplectic.
\end{enumerate}
\end{proposition}
Note that it follows from the first property that
\beq \lab{f-d}
\ell(-d) = \log \E |\det A |^{-1}
\eeq
which extends the well-known observation that \cite{furs63,zeld-et-al}
\beq \lab{f-d0}
\ell(-d)=0 \inter{if} \det A =1.
\eeq

The properties in proposition \ref{prop:elltm} are established in Appendix \ref{app:prop}. They are useful in practice: because numerical methods for the estimation of the Lyapunov exponents are more accurate when $|q|$ is small, estimates for $\ell(q)$ with $q<-d$ can be obtained efficiently by evaluating $\tilde \ell(-q-d)$. Also, the replica method (described in section \ref{rep}), which provides estimates of $\ell(q)$ for $q$ even and positive, can be used for some negative values of $q$ when the proposition \ref{prop:elltm} is exploited.

As a practical tool for the computation of generalised Lyapunov exponents, the eigenvalue problem \eqn{eig} appears limited to small matrices with $d=2$ or $d=3$, because it requires the discretisation of an operator acting on functions of $d-1$ variables. (See Refs. \cite{hayn-v05,tzel-hayn10} and section \ref{soleig} below for some implementations with $d=2$.) In the next section we describe a Monte Carlo method that does not suffer from this limitation.

\section{Resampled Monte Carlo}  \label{rmc}

The simplest Monte Carlo method for the estimation of $\ell(q)$, which we term 
brute-force Monte Carlo, consists in computing the estimator 
\[
\Zbf_N = \frac{1}{K} \sum_{k=1}^K \|A_N\su{k} \cdots A_1\su{k} x_0\|^q,
\]
where the bracketed superscript indexes $K$ independent realisations of the sequences of random matrices $A_n$. Clearly,
\[
\E \Zbf_N \asymp \e^{N \ell(q)},
\]
so $N^{-1} \log \Zbf_N$ estimates $\ell(q)$. This method is hopelessly inefficient, however, unless $|q|$ is small. To see why, note that the variance is of $\Zbf_N$ is given by
\[
\var \Zbf_N = \frac{1}{K} \var \| X_N \|^q = \frac{1}{K} \left( \E \| X_N \|^{2q} - (\E \| X_N \|)^q)^2 \right) \sim \frac{c_{2q} \e^{N\ell(2q)}- c_q \e^{2N\ell(q)}}{K}.
\]
The convexity of $\ell(q)$ then implies that $\exp(N \ell(2q)) \gg \exp(2N \ell(q))$.
So
the variance of $\Zbf_N$ is exponentially large in $N$, and a number of realisations $K \gg \exp[N(\ell(2q)-2\ell(q))]$ is in principle necessary for an accurate estimation of $\ell(q)$. 

The inefficiency of the brute-force Monte Carlo estimate stems from the fact that for finite $q$, $\E \| X_N \|^q$ is dominated by rare realisations which are undersampled unless $K$ is exponentially large. To remedy this, we can resample at each iteration so that the dominant contributions to $\E \| X_N \|^q$ are represented by more realisations; this is the main idea behind sequential importance sampling or `go-with-the-winners' strategies \citep{liu01,gras02}. We describe a particularly simple algorithm for such a resampling strategy which we term Resampled Monte Carlo (RMC).

\subsection{Algorithm} \label{algo}

Like the brute-force Monte Carlo, the algorithm relies on $N$ iterations and $K$ realisations, calculating $X\su{k}_n$ for $n=1,\cdots,N$ and $k=1,\cdots,K$. The difference is that the realisations are dependent. Rather than $X\su{k}_n$, it is convenient to use the corresponding unit vector
\[
E\su{k}_n = \frac{X\su{k}_n}{\| X\su{k}_n\|}.
\] 
Starting with $E\su{k}_0=x_0$ for $k=1,\cdots,K$, the algorithm proceeds iteratively with two steps at each iteration $n$:
\begin{enumerate}
\item Draw $K$ random matrices $A\su{k}_n$, and compute
\beq \lab{algo1}
\hat E\su{k}_n = \frac{A\su{k}_n E\su{k}_{n-1}}{\|A\su{k}_n E\su{k}_{n-1}\|} \inter{and} \alpha\su{k}_n = \|A\su{k}_n E\su{k}_{n-1}\|^q.
\eeq
\item Resample by letting
\beq \lab{algo2}
E\su{k}_n = \hat E\su{J\su{k}_n}_n.
\eeq
Here the $J\su{k}_n$ are independent random variables taking values in $\{1,\cdots,K\}$, with
\beq \lab{algo3}
\P\left( J\su{k}_n = j\right) = \frac{\alpha\su{j}_n}{\beta_n}, \inter{where} 
\beta_n = \sum_{k=1}^K \alpha\su{k}_n.
\eeq
\end{enumerate}
The estimate of $\E \|X_N\|^q$ is then given by
\beq \lab{Zn}
Z_N = \frac{1}{K^N} \beta_1 \beta_2 \cdots \beta_N.
\eeq

Note that the resampling step ensures that, at each iteration $n$, the weight of each realisation in the estimate of $\E \| X_n \|^q$ is the same. Note also that the resampling is tailored to a specific value of $q$. Unlike in the brute-force Monte Carlo, where the same ensemble can be used to estimate $\ell(q)$ for a range of values of $q$, the RMC approach requires a new sampling for each value of $q$ (although it may be possible to use the same sampling for a narrow enough range of $q$). In several applications, though, $\ell(q)$ is only required for a single value of $q$ \citep[e.g.][]{anto-et-al,tsan-et-al05a,hayn-v05}. 

In Appendix \ref{pseudo} we give a pseudocode for the RMC algorithm. This  illustrates the simplicity of the algorithm, and should be useful for readers wishing to implement it in a specific programming language.

\subsection{Analysis} \lab{anal}

To analyse the algorithm further, we note that the $NK$ random matrices $A\su{k}_n$ involved in the computation form $K$ independent paths consisting of the $N$ matrices that are multiplied in succession to obtain each $E\su{k}_N$.  These paths are
\[
A\su{I_N\su{k}}_N,\, A\su{I\su{k}_{N-1}}_{N-1},\, \cdots,\, A\su{I\su{k}_1}_1,
\] 
where the random variables $I\su{k}_n, \, n=1,\cdots,N$ are determined by $k$ and by the random variables $J\su{k}_n$ according to
\[
I\su{k}_{N}=J\su{k}_{N}, \ \  I\su{k}_{N-1}=J\su{I\su{k}_N}_{N-1},  \ \ I\su{k}_{N-2}=J\su{I\su{k}_{N-1}}_{N-2}, \cdots.
\]
The factors $\alpha_n\su{\cdot}$ that are computed along the path that yields $E\su{k}_N$ are then
\beq \lab{alphaIk}
\alpha\su{I\su{k}_n}_n = \frac{\|A\su{I\su{k}_n}_n A\su{I\su{k}_{n-1}}_{n-1} \cdots A\su{I\su{k}_1}_1 x_0\|^q}{ \|A\su{I\su{k}_{n-1}}_{n-1} A\su{I\su{k}_{n-2}}_{n-2} \cdots A\su{I\su{k}_1}_1 x_0 \|^q}, \ n=1,\cdots,N.
\eeq
Note that the distribution of the $I\su{k}_n$ is that same as that of the $J\su{k}_n$, since the distribution of the latter is independent of $k$; thus,
\[
\P\left( I\su{k}_n = j\right) = \frac{\alpha\su{j}_n}{\beta_n}.
\]
For a given realisation of the matrices $A_n\su{k}$ for $n=1,\cdots,N$ and $k=1,\cdots,K$, the probability of a particular path
\[
A\su{j_N}_N,\, A\su{j_{N-1}}_{N-1},\, \cdots A_1\su{j_1}
\]
is then
\beq \lab{pathprob}
\P(j_1,\cdots,j_N| A_n\su{k})=
 \frac{\alpha_1\su{j_1} \alpha_2\su{j_2} \cdots \alpha_{N}\su{j_{N}} }{\beta_{1} \beta_{2} \cdots \beta_N},
\eeq
where
\beq \lab{alphajn}
\alpha_n\su{j_n} = \frac{\|A_n\su{j_n} \cdots A_1\su{j_1} x_0\|^q}{\|A_{n-1}\su{j_{n-1}} \cdots A_1\su{j_1} x_0\|^q} \inter{and} \beta_n = \sum_{j_n=1}^K \alpha\su{j_n}_n.
\eeq
Here we abuse notation slightly and use the same symbol $\alpha$ to denote, in \eqn{alphaIk}, a random variable that depend on both the $A\su{k}_n$ and the $J\su{k}_n$, and in \eqn{alphajn} one that depend only on the  $A\su{k}_n$; the same abuse of notation is made for $\beta$.

To compute the expected value of functions $f$ of $Z_N$ produced by the algorithm, we note that the corresponding expectation, $\E'$ is a combination of the expectation $\E$ over the random matrices $A\su{k}_n$ and of the expectation over the random variables $J\su{k}_n$. Using \eqn{pathprob} to compute the latter expectation leads to
\beq \lab{Ea}
 \E' f(Z_N) = \E \sum_{j_1,\cdots, j_N=1}^K \frac{\alpha_1\su{j_1} \alpha_2\su{j_2} \cdots \alpha_{N}\su{j_{N}} }{\beta_{1} \beta_{2} \cdots \beta_N} f(z_N),
 \eeq
where $z_N = \beta_1 \beta_2 \cdots \beta_N/K^N$, and the $\alpha\su{j_n}_n$ and $\beta_n$ defined as in \eqn{alphajn}. 
 
Using \eqn{Ea}, it is immediate to establish 
\begin{proposition}
$Z_N$ is an unbiased estimator for $\E \| X_N \|^q$: 
\beq \lab{eZN}
\E' Z_N = \E \| X_N \|^q.
\eeq
\end{proposition}
This follows from the computation
\begin{eqnarray*}
\E' Z_N &=& \frac{1}{K^N} \E \, \sum_{j_1,\cdots, j_N=1}^K \alpha_1\su{j_1} \alpha_2\su{j_2} \cdots \alpha_{N}\su{j_{N}} 
=  \frac{1}{K^N} \E \, \sum_{j_1,\cdots, j_N=1}^K 
\|A_N\su{j_N} \cdots A_1\su{j_1} x_0\|^q \\
&=& \E \|A_N \cdots A_1 x_0\|^q = \E \| X_N \|^q,
\end{eqnarray*}
which uses \eqn{Zn}, \eqn{alphajn} and \eqn{Ea}. 

In order to estimate the error of $Z_N$, we obtain the following expression:
\begin{proposition} The expected value of $Z_N^2$ is
\beq \lab{eZN2}
\E' Z_N^2 = \frac{1}{K^{2N}} \, \E \sum_{j_1,\cdots, j_N=1}^K \sum_{j'_1,\cdots, j'_N=1}^K \|A_N\su{j_N} \cdots A_1\su{j_1} x_0\|^q \|A_N\su{j'_N} \cdots A_1\su{j'_1} x_0\|^q.
\eeq
\end{proposition}
This is obtained from
\[
\E' Z_N^2 = \frac{1}{K^{2N}} \, \E \sum_{j_1,\cdots, j_N=1}^K 
{\alpha_1\su{j_1} \alpha_2\su{j_2} \cdots \alpha_{N}\su{j_{N}} }{\beta_{1} \beta_{2} \cdots \beta_N} 
\]
on using the definition \eqn{alphajn} of the $\beta_n$. 

Expression \eqn{eZN2} makes clear why the variance of  $Z_N$ is much smaller than that of $Z_N^\mathrm{bf}$. Only $K^N$ terms of the $K^{2N}$ terms in \eqn{eZN2} lead to contributions proportional to $\exp(N \ell(2q))$ (those for which $j_1=j_1', \cdots j_N=j_N'$) with all the others leading to much smaller contributions with, in particular, $(K(K-1))^N$ proportional to $\exp(N \ell(q))$  (those for which $j_1\not=j_1', \cdots j_N\not=j_N'$). In contrast, in $\E (Z_N^\mathrm{bf})^2$, all the terms are proportional to $\exp(N \ell(2q))$. 

The improvement can be evaluated explicitly in the scalar case $d=1$. Admittedly, this is an uninteresting case as far as the numerical evaluation of $\ell(q)$ is concerned, since \eqn{glyapdef} holds exactly for finite $N$, but it is instructive nonetheless. For $d=1$, the asymptotic relation \eqn{asymEXN} holds exactly for all $N$ and with $c_q$=1. It follows that the terms in \eqn{eZN2} can be evaluated explicitly : if $j_k=j_k'$ for $l$ values of $k$ and $j_k\not=j_k'$ for the remaining $N-l$ values, 
\[
\E  \|A_N\su{j_N} \cdots A_1\su{j_1} x_0\|^q \|A_N\su{j'_N} \cdots A_1\su{j'_1} x_0\|^q = \e^{l \ell(2q) + 2 (N-l) \ell(q)}
\]
Since there are ${N \choose l} K^N (K-1)^{N-l}$ such terms, \eqn{eZN2} becomes
\[
\E' Z_N^2 = \frac{1}{K^N} \sum_{l=0}^N {N \choose l} (K-1)^{N-l} \e^{l \ell(2q) + 2 (N-l) \ell(q)} = \frac{1}{K^N} \left(\e^{\ell(2q)} + (K-1) \e^{2 \ell(q)} \right)^N
\]
The variance is then
\[
\var Z_N = \frac{\left(\e^{2 \ell(q)} (K + \gamma_q) \right)^N-K^N \e^{2N \ell(q)}}{K^N} 
\sim \frac{N}{K} \gamma_q \e^{2N \ell(q)},
\]
where we have introduced 
\beq \lab{gammaq}
\gamma_q = \e^{\ell(2q)-2\ell(q)}-1. 
\eeq
Thus the relative variance of $Z_N$ is 
\beq \lab{relvar}
\frac{\var Z_N}{\left( \E Z_N \right)^2} \sim \frac{N \gamma_q}{K}, 
\eeq
and the Monte Carlo estimation of $\E \| X_N \|^q$ by ${Z_N}$ requires only that $K \gg N$ rather than $K \gg \exp [N (\ell(2q)-\ell(q))]$ as is the case for the brute-force Monte Carlo.  This drastic gain in computational efficiency is expected to apply also for matrices with $d>1$: the non-commutativity is likely to modify \eqn{relvar} only through the introduction of an $N$- and $K$-independent factor on the right-hand side.

Although we have found that the RMC algorithm performs very well for a broad range of random-matrix products, it is useful to have alternatives methods of evaluating $\ell(q)$ at one's disposal. This provides independent checks for the RMC results or, in the case of asymptotic approximation for $|q| \gg 1$, makes it possible to estimate $\ell(q)$ when the RMC approach becomes unreliable. Such alternative methods are discussed in the next section.

\section{Other estimates} \label{other}

\subsection{Solving the eigenvalue problem \eqn{eig}} \label{soleig}

For $d=2$ or $3$, it is practical to compute $\ell(q)$ as the largest eigenvalue of the eigenvalue problem \eqn{eig} for  functions $v$ on $S^{d-1}$. Here we describe an implementation for $d=2$. In this case, the unit vector $\n$ can be parameterised by an angle $\theta$, and $v$ can be expanded in a Fourier series, which we write as
\[
v(\theta) = \Re \sum_{n=0}^{M-1} v_n \e^{\i n \theta},
\]
and truncate at some $M$. A straighforward discretisation of the eigenvalue problem \eqn{eig} is then obtained by collocation at points $\theta_m = 2 \pi m/M,\, m=0,\cdots,M-1$. This leads to the generalised matrix eigenvalue problem
\beq \lab{PQeig}
P \bv = \lambda Q \bv, 
\eeq
where $\bv=(v_0,\cdots,v_{M-1})^\mathrm{T}$, and the $M \times M$ matrices $P$ and $Q$ have entries given by
\beq \lab{PQ}
P_{mn} = \E \| A \n(\theta_m)\|^q \e^{\i n \Theta(\theta_m)} \inter{and}
Q_{mn} = \e^{\i n \theta_m},
\eeq
where $\n(\theta_m)=(\cos \theta_m, \sin \theta_m)^\mathrm{T}$ and $\Theta(\theta_m)$ is defined by
\beq \lab{Theta}
\left( \begin{array}{c}
\cos \Theta(\theta_m) \\ \sin \Theta(\theta_m) \end{array}
\right) = \frac{A \n(\theta_m)}{\|A \n(\theta_m)\|}.
\eeq
The expectation in the definition of $P$ can be computed using a Monte Carlo approach, and the eigenvalue problem solved using a standard technique. 

\subsection{Replica method for positive even $q$} \label{rep}

A useful method, known as the replica trick \citep[see][and reference therein]{cris-et-al93}, makes it possible to compute $\ell(q)$ for $q$ positive and even by finding the largest eigenvalue of a $qd \times qd$ (deterministic) matrix. To see how this can be achieved, observe that the $q$-fold tensor product $X_n$ with itself satisfies
\beq
X_n^{\otimes q} = A_n ^{\otimes q} X_{n-1}^{\otimes q},
\eeq
where $A_n ^{\otimes q}$ is the $q$-fold Kronecker product of $A_n$ with itself.
Taking the expectation then leads to
\beq
\E X_n^{\otimes q} = \E A_n ^{\otimes q} \E X_{n-1}^{\otimes q}.
\eeq
Therefore
\beq
 \E X_N^{\otimes q} \asymp \e^{\mu_q N} y,
\eeq
where  $\mu_q$ is the largest eigenvalue of the  $qd \times qd$ matrix $ \E A_n ^{\otimes q}$, and $y \in \mathbb{R}^{qd}$ is the corresponding eigenvector.  Since for $q>0$ even, $\|X_N\|^q$ is obtained from $X_N^{\otimes q}$ by contraction, 
\beq
\ell(q) = \mu_q \inter{for} q > 0 \ \ \textrm{even}.
\eeq
The results extends to the case of odd $q$ when the matrices $A$ have  only non-negative entries.

\subsection{Large-$|q|$ asymptotics} \label{asy}

For large $|q|$, numerical methods that involve taking expectations by sampling become inefficient, and it is useful to develop analytic or semi-analytic methods that take advantage of $|q| \gg 1$ to provide an asymptotic estimate for $\ell(q)$. The eigenvalue problem \eqn{eig} is a good starting point. Since the expectation is an integral over the random parameters that define the matrix ensemble, we can attempt to approximate this integral for $|q| \gg 1$ using Laplace's method. A dominant-balance argument suggests that the eigenfunction $v(\n)$, which depends implicitly on $q$, should have the asymptotic WKB form
\beq
v(\n) \sim z(\n) \e^{q w(\n)},
\eeq
where $w$ and $z$ are independent of $q$. Substituting this into \eqn{eig} gives
\beq \lab{zew}
\lambda z(\n) \e^{q w(\n)} = \E \e^{q \left(\log \|A \n \| + w(A \n/\|A \n\|) \right)} z(A \n/\|A \n\|).
\eeq

When the values of $\| A \|$ are bounded, the expectation on the right-hand side is dominated by the matrices maximising the argument of the exponential (assuming a non-zero probability density for the maximising matrices). Concentrating on the case $q>0$, this gives
\beq \lab{eigw}
w(\n) = \sup_A \left( \log \|A \n \| + w(A \n/\|A \n\|)  \right) - \kappa,
\eeq
for some constant $\kappa$, where the supremum is over the support of the probability measure of the random matrices. Note that $w$ is defined up to the addition of an arbitrary constant. Equation \eqn{eigw} can be interpreted as a nonlinear eigenvalue problem, with $w$ as the eigenfunction and $\kappa$ as the eigenvalue. If this eigenvalue problem has a solution, the largest value of $\kappa$ governs the rough asymptotics of $\lambda_1$ and hence the asymptotics of $\ell(q)$, with the result
\beq \lab{ellas}
\ell(q) \sim \kappa q.
\eeq 
Note that this behaviour implies that the rate function $g(h)$ of the finite-time Lyapunov exponents has a vertical asymptote for $h=\kappa$. Therefore $\kappa$ is also given by the maximum possible (largest) finite-time Lyapunov exponent:
\beq
\kappa =  \lim_{N \to \infty} \sup_{A_1 \cdots A_N} \frac{1}{N} \log \|A_N \cdots A_1\| .
\eeq
It would of course be difficult to attempt to determine $\kappa$ by sampling the right-hand side of this expression. In general, $\kappa \le \sup_A \log \| A \|$, with the equality holding only in special cases; see Appendix \ref{bound}.

The result \eqn{ellas} can be refined by noting that Laplace's method applied to \eqn{zew} leads to the expectation of a term of the form $\exp(-q \langle A-A_*,A-A_* \rangle)$, where $A_*$ is the maximiser in \eqn{eigw} and $\langle \cdot, \cdot \rangle$ is some scalar product (both $A_*$ and $\langle \cdot, \cdot \rangle$ depend on $\n$). Carrying out the expectation yields a factor $q^{-D/2}$, where $D$ is the dimension of the support of the measure. It follows that
\beq \lab{as1}
\ell(q) \sim \kappa q - \frac{D}{2} \log q + O(1).
\eeq
This asymptotics implies that $g \sim  - D\log(\kappa-h)/2$, which describes  the manner in which $g(h)$ approaches the vertical asymptote at $k=\kappa$.

When $\|A\|$ is unbounded, the matrices $A$ dominating the expectation in \eqn{zew} are controlled by a balance between the argument of the exponential, which grows with $\|A\|$, and the probability density of $A$ which should  
decrease with $\|A\|$ if $\ell(q)$ is to be finite. This means that one needs to apply Laplace's method for movable maxima \citep[e.g.][]{bend-orsz} and consider the $q$-dependent maximum of
\beq
\log \|A \n \| + w(A \n/\|A \n\|) + q^{-1} \log \pi(A),
\eeq
where $\pi(A)$ is the probability density of $A$ and $w$ depends on $q$. For instance, if $\pi(A)$ is Gaussian, this maximum corresponds to matrices $A$ with $O(q^{1/2})$ entries, leading without further calculations to
\beq \lab{gaussian}
\lambda \asymp \e^{q \log q^{1/2}}, \inter{hence}
\ell(q) \sim \frac{q}{2} \log q.
\eeq
Correspondingly, $g(h) \asymp \e^{h}$ for $h \gg 1$.

\section{Examples} \label{ex}

\subsection{Two-dimensional sine map}

In studies of transport and mixing by complex fluid flows, numerous authors have used the random sine map proposed by Pierrehumbert \cite{pier94} as a model of a completely chaotic, non-divergent flow. In two dimensions, this map is given by 
\beq \lab{sin2d}
x_{n+1} = x_n + a \sin(y_n + \phi_1), \quad
y_{n+1} = y_n + b \sin(x_{n+1} + \phi_2),
\eeq
where $a$ and $b$ are fixed parameters, and the random angle $\phi_1$ and $\phi_2$ are independent and uniformly distributed in $[0, 2 \pi]$. The Jacobian matrix $\partial(x_{n+1}, y_{n+1})/\partial(x_n,y_n)$, whose statistics are independent of $(x_n,y_n)$, is given at $(0,0)$ by
\beq \lab{sine2d}
A = \left(
   \begin{array}{cc} 
      1 & a \cos \phi_1\\
      b \cos(\sin \phi_1 + \phi_2) & 1 + ab  \cos \phi_1 \cos(\sin \phi_1 + \phi_2). \\
   \end{array} \right)
\eeq
It satisfies $\det A =1$ and hence, since $d=2$, is symplectic.

The generalised Lyapunov exponents corresponding to the ensemble of matrices $A$ generated by $\phi_1$ and $\phi_2$ characterise the separation of nearby particle in the sine flow. Remarkably, their knowledge makes it possible to predict, in some cases at least, the rate of decrease of the variance of a passive scalar released in the flow \citep{balk-foux,anto-et-al,tsan-et-al05b,hayn-v05}. Specifically, this rate is given by $g(0)$ and, in view of the Legendre duality of $g(h)$ and $\ell(q)$, by $-\ell(q_*)$, where $q_*$ is such that $\ell'(q_*)=0$. Because of property 3 of proposition \ref{prop:elltm}, $q_*=-1$. 

\begin{figure}
\begin{center}
\includegraphics[height=7cm]{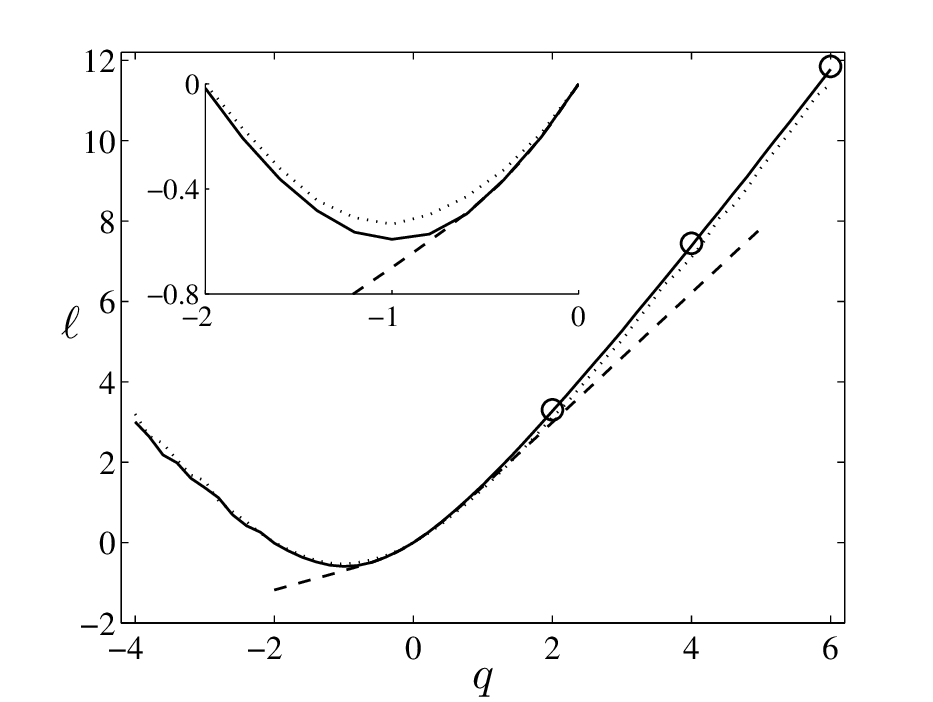}
\caption{Generalised Lyapunov exponents of the product of the random matrices \eqn{sine2d}  with $a=b=\pi$. The result of the RMC method (solid line) are compared with those of the brute-force Monte Carlo method (dashed line), and of the numerical solution of the eigenvalue problem (dotted line). The inset displays a close up of the region $-2 \le q \le 0$.}
\label{fig:sine2gen}
\end{center}
\end{figure}

In the literature, $\ell(-1)$ has been evaluated using brute force Monte Carlo \cite{fere-hayn,tsan-et-al05a} and solving the eigenvalue problem \cite{hayn-v05,tzel-hayn10}. Here we apply the algorithm of section \ref{rmc} to demonstrate its efficiency. In Figure \ref{fig:sine2gen}, we compare $\ell(q)$ obtained for $a=b=\pi$ using different numerical methods: brute force Monte Carlo, RMC, and numerical solution of the eigenvalue problem using 128 Fourier modes and 128 collocation points. For the latter two methods, we have used a relatively small ensemble, with $K=1000$, while for the brute force computation we have used the much larger $K=10^5$. The number of matrix multiplication $N$ was taken as 100 for the RMC but only $N=50$ for the brute-force computation which is restricted  to moderate values of $N$.
Also shown are the very reliable estimates obtained for $q=2,\, 4$ and $6$ using the replica method. The figure illustrates how impractical the brute force computation is to estimate $\ell(q)$ for, say, $q>2$ and $q<0$. The other methods, by contrast, provide good estimates for a wide range of $q$. Based on the comparison with the replica estimate, the RMC algorithm, which for the parameters chosen is the faster by a factor of about 5, appears to be the more accurate method. The inset in the figure zooms on the range $q \in [-2,0]$ to emphasise the substantial differences in the estimates in that region leading, in particular, the inaccuracy in the estimates of $\ell(-1)$ needed for decay-rate predictions in the passive-scalar problem. In this regard, we note that a sequence of 500 RMC computations gives the average and standard deviation $\ell(-1)=-0.5916 \pm 0.0056$.

\begin{table}
\begin{center}
\begin{tabular}{r|cccc}
\hline
$K=$ & &500 & 1000 & 2000 \\
 \hline
$N=20$  & & 0.12 & 0.052 & 0.027 \\
40  & & 0.24 & 0.12  & 0.059 \\
80 & & 0.62 & 0.23 & 0.10 \\
\hline
\end{tabular}
\caption{Estimation of the normalised variance $\var Z_N/(\E Z_N)^2$ for   the RMC method applied to the matrices \eqn{sine2d} for $q=-1$ and for different values of the number of realisations $K$ and number of iterations $N$.} \label{tab:var}
\end{center}
\end{table}

We have used the example of the two-dimensional sine flow with $q=-1$ to assess 
the dependence of the variance of the RMC estimate $Z_N$ on the number of realisations $K$ and on the number of iterations $N$. We have estimated this variance by performing 500 computations of $Z_N$ for 9 combinations of the parameters $K$ and $N$. The results are reported in Table \ref{tab:var}. Unsurprisingly, the sample variance scales roughly like $1/K$; more interestingly it also scales like $N$ in agreement with the behaviour \eqn{relvar} obtained in the scalar case. The behaviour \eqn{relvar} can be tested further:  since $\ell(-2)=0$, $\gamma_{-1} = \exp(-2 \ell(-1))-1 \approx 2.26$, which compares reasonably well with the various estimates of $\var Z_N/(\E Z_N)^2$ that can be obtained from table \ref{tab:var}. 
 
Returning to figure \ref{fig:sine2gen}, we note that the estimates of $\ell(q)$ appear less accurate for negative $q$ unless $|q| \lesssim 1$; this can easily be remedied, however, by using the third property in proposition \ref{prop:elltm}, namely $\ell(q)=\ell(-q-2)$, so that the only negative range that needs to be considered is $q \in [-1,0]$. 

\begin{figure}
\begin{center}
\includegraphics[height=7cm]{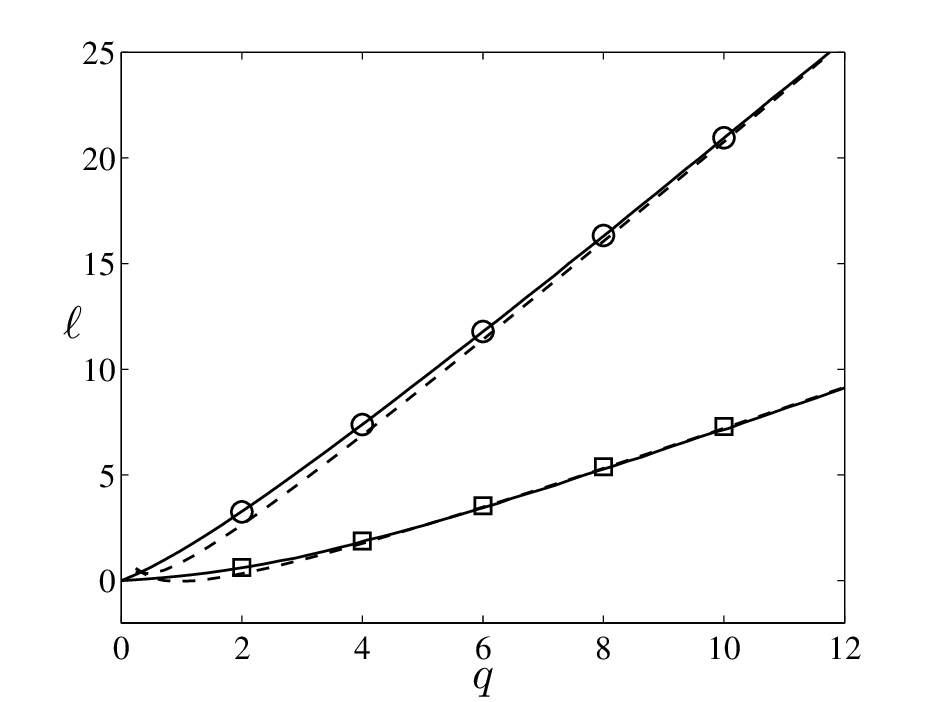}
\caption{Generalised Lyapunov exponents of the product of the random matrices \eqn{sine2d}  with $a=b=\pi$ (circles), and $a=\pi$ and $b=\pi/8$ (squares). The result of the RMC method (solid lines) are compared with the large-$q$ asymptotic estimate (dashed lines), and the results of the replica method (symbols).}
\label{fig:sine2dasy}
\end{center}
\end{figure}

The estimation of $\ell(q)$ is truly challenging for large $q$. Here, we briefly consider it for the matrices \eqn{sine2d} in order to assess both the reliability of the RMC method, and the asymptotic estimate \eqn{as1}. 
Figure \ref{fig:sine2dasy} shows $\ell(q)$ for the matrices \eqn{sine2d} with $a=b=\pi$, and with $a=\pi$ and $b=\pi/8$. In both cases, $\ell(q)$ can be approximated according to \eqn{as1} with $D=2$ (since the matrices are defined by 2 random angles $\phi_1$ and $\phi_2$). The value  of $\kappa$ should be derived by solving \eqn{eigw}. The case $a=b$ is special, however. It can be verified in this case that the maximum of $\| A \n \|$
is achieved for matrices $A$ and unit vectors $\n$  such that $A \n /\|A \n\| = \n$. As a consequence, we have that
\beq \lab{kappamax}
\kappa =  \log  \sup_{A} \|A \| \inter{\textrm{for}} a=b.
\eeq
This result, which holds for any matrix ensemble such that $A \n /\|A \n\| = \n$ for $A$ and $\n$ maximizing $\| A \n \|$, is established in Appendix \ref{bound}. It enables a simple evaluation of $\kappa$ when $a=b$, giving $\kappa=2.467$. 
The corresponding asymptotic estimate \eqn{as1} is compared in Figure \ref{fig:sine2dasy} with the numerical estimates obtained using the RMC and replica methods. The $O(1)$ term in \eqn{as1} is determined by matching the asymptotic and numerical results for the largest value of $q$ on the figure. The figure demonstrates the validity of the asymptotic estimate; it also illustrates the reliability of the RMC method (used here with an ensemble of $K=1000$ matrices) which provides accurate estimates of $\ell(q)$ for $q$ as large as 12, at least for matrices considered here.

\begin{figure}
\begin{center}
\includegraphics[height=7cm]{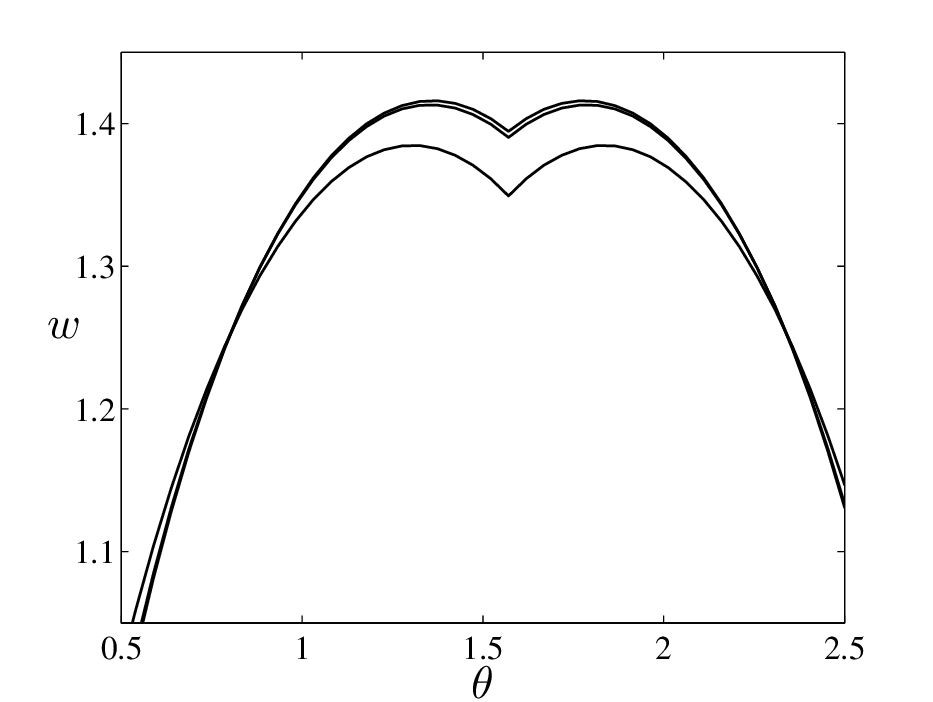}
\caption{Iterates $w\su{k}(\theta)$ for $k=1,2,3$ in the numerical solution of the problem \eqn{eigw} determining $\kappa$ for the matrices \eqn{sine2d} with $a=\pi$ and $b=\pi/8$.}
\label{fig:iterasy}
\end{center}
\end{figure}

The simple result \eqn{kappamax} is very special. In general, when $a \not= b$, the right-hand side of \eqn{kappamax} is a strict upper bound for $\kappa$. There is then no explicit expression for $\kappa$, and the problem \eqn{eigw} must be solved for both $\kappa$ and $w(\n)$. We have implemented a numerical solution of this problem for the matrices \eqn{sine2d}. The implementation relies on an iteration: successive iterates $w\su{k}, \, k=1,2,\cdots$, regarded as functions of the angle $\theta$ parameterising $\n$,  are represented using the truncated Fourier series
\[
w\su{k}(\theta)=\Re \sum_{n=1}^{M-1} w_n\su{k} \e^{\i n \theta},
\]
from which the average ($n=0$) term is omitted in order to fix the arbitrary constant in the definition of $w$. The iteration scheme
\beq \lab{iter}
w\su{k+1}(\theta_m) + \kappa\su{k+1} = \sup_A \left(\log \| A \n(\theta_m) \| + \sum_{n=1}^{M-1} w_n\su{k} \e^{\i n \Theta(\theta_m)}\right),
\eeq
where $\Theta(\theta_m)$ is defined in \eqn{Theta},
determines $w^{(k+1)}$ on the grid points $\theta_m=2 m\pi/M$, with $\kappa\su{k+1}$ fixed using the condition of zero average for $w\su{k+1}$. The supremum is evaluated numerically by finding the maximum over a finite number of matrices $A$ obtained for values of $\phi_1$ and $\phi_2$ on a grid. An inverse Fourier transform then gives $w_n\su{k+1}$, and the iteration can continue. Figure \ref{fig:iterasy} shows the first three iterates of this method applied in the case $a=\pi$ and $b=\pi/8$. The functions $w\su{k}(\theta)$ are defined for $\theta=[0,2\pi]$ and $\pi$-periodic; here we show an interval of $\theta$ around the maxima of these functions. The first iterate, corresponding to the lowest curve, is simply $w\su{1}(\theta)=\sup_A \log  \| A (\n(\theta)\|$. The next two iterates illustrate the rapid convergence of the method; after 4 iterations, convergence is achieved, and the estimate $\kappa=1.061$ is obtained; this is substantially less than $ \log \sup_{A} \| A \| = 1.385$. The validity of our asymptotic formula and evaluation of $\kappa$ are confirmed by Figure \ref{fig:sine2dasy} which shows an excellent match between the asymptotic and numerical estimates of $\ell(q)$. A similar match was found for other values of $a$ and $b$.

\subsection{Three-dimensional sine map}

In order to explore matrices that are not symplectic but have determinant 1, we consider the stretching by the volume-preserving map of $\mathbb{R}^3$
\beq \lab{sin3d}
x_{n+1} = x_n + a \sin(y_n + \phi_1), \quad 
y_{n+1} = y_n + b \sin(z_n + \phi_2), \quad
z_{n+1} = z_n + c \sin(x_{n+1} + \phi_3),
\eeq
where the $\phi_j, \, j=1,2,3$ are independent uniformly distributed in $[0,2\pi]$. This map generalises to three dimensions the two-dimensional alternating sine map \eqn{sin2d}.
The corresponding Jacobian matrix at the origin is
\beq \lab{sine3d}
A = \left( \begin{array}{ccc}
1 & a \cos \phi_1 & 0 \\
0 & 1 & b \cos \phi_2 \\
c \cos (a \sin\phi_1 +\phi_3) & a c  \cos \phi_1  \cos (a \sin\phi_1 +\phi_3) 
& 1 \end{array} \right).
\eeq 

\begin{figure}
\begin{center}
\includegraphics[height=7cm]{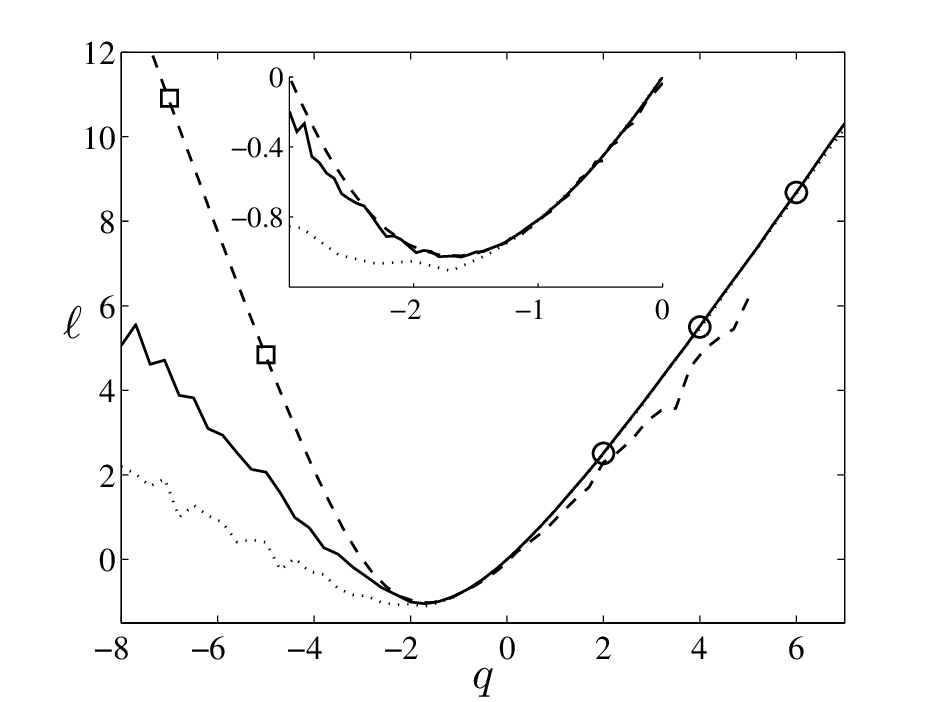}
\caption{Generalised Lyapunov exponents of the product of the random matrices \eqn{sine3d}  with $a=b=c=\pi$. The RMC estimates of $\ell(q)$ obtained for $N=100$ with an ensemble size $K=1000$ (solid line) and $K=100$ (dotted line) are compared  with an estimate of $\ell^{-}(-q-3)$ obtained by applying the RMC method to $A^{-1}$ with $K=1000$ (dashed line). The results of the replica method, applied to $A$ (circles) and $A^{-1}$ (squares) are also indicated. The curves in the inset, which displays a close up of the region $-3 \le q \le 0$, have been computed using $K=5000$. }
\label{fig:sine3gen}
\end{center}
\end{figure}

The results of several numerical computations with these matrices are displayed in Figure \ref{fig:sine3gen}. In the main panel, we show three different estimates of $\ell(q)$, all obtained using the RMC method with $N=100$. The first (solid line) applies the RMC method to the matrices $A$ with an ensemble size $K=1000$; the second (dotted line) also uses the RMC method but with the much smaller ensemble size $K=100$.  
The results illustrate the difficulties that arise when evaluating numerically $\ell(q)$ for $q<0$: for $q \lesssim -2$ in this case, the numerical estimates appear to be very unreliable, and the situation does not improve much when the number of realisations is increased from $K=100$ to $K=1000$. The problem is easily remedied, however, using property 2 of proposition \ref{prop:elltm}: by applying the RMC algorithm to $A^{-1}$ rather than to $A$, we estimate $\ell^{-}(q)$; this estimate, which proves accurate for $q \gtrsim -2$, then provides a reliable approximation for $\ell(q)$ with $q\lesssim-1$ since $\ell(q)=\ell^-(-q-3)$. The curve of $\ell^{-}(-q-3)$ is shown by the dashed curve in Figure \ref{fig:sine3gen}. The best estimate of $\ell(q)$ should be read as the dashed curved for $q \lesssim-1$ and the solid curve for $q \gtrsim -2$. For definiteness, one could choose the point $q=-d/2=-3/2$ for the transition  between the two approximations. 


\subsection{Gaussian matrices}

\begin{figure}
\begin{center}
\includegraphics[height=7cm]{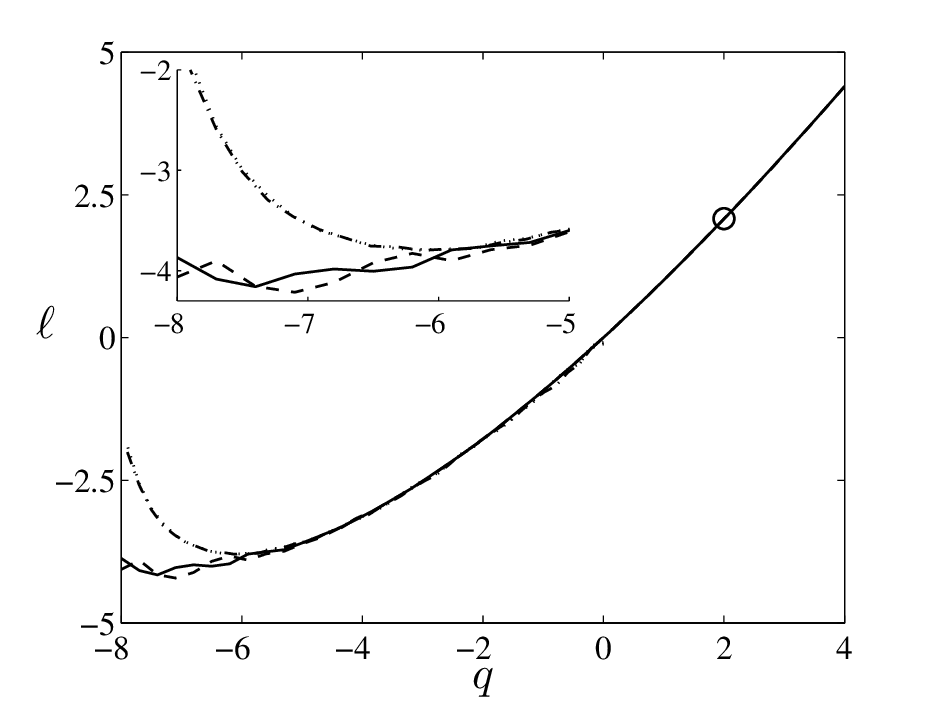}
\caption{Generalised Lyapunov exponents of the product of $8 \times 8$ symmetric matrices with independent, zero-mean and variance-one Gaussian upper-diagonal entries. The RMC estimates for $\ell(q)$ obtained with $K=2000$ (dashed line) and $K=4000$ (solid line) realisations are compared with the RMC estimates for $\tilde \ell(-q-8)$ obtained with $K=2000$ (dash-dotted line) and $K=4000$ (dotted line, almost indistinguishable from the dash-dotted line). The replica estimate for $q=2$ is indicated by the circle.}
\label{fig:gauss}
\end{center}
\end{figure}

As a last example, we consider the case of Gaussian matrices. When all the entries are independent $N(0,\sigma^2)$ variables, the statistics of $\| A \n \|$ are independent of $\n$, which leads to the explicit expression
\beq \lab{gausRMCo}
\ell(q)=\log \E \|A \n \|^q = \frac{q}{2} \log (2  \sigma^2) + \log \Gamma \left(\frac{q+d}{2} \right) - \log \Gamma\left(\frac{d}{2}\right),
\eeq
for $q > -d$, with $\ell(q)=\infty$ for $q \le -d$ \citep{cris-et-al93}.  No such explicit expressions are available when the entries are correlated, however, and $\ell(q)$ needs to be estimated numerically. Here, we examine the case of symmetric matrices with iid $N(0,\sigma^2)$ upper-diagonal entries. As in the case of independent entries, $\ell(q)=\infty$ for $q<-d$, and so we can expect  difficulties in estimating $\ell(q)$ for values of $q$ slightly larger than $-d$, say for $q \lesssim -d/2$. It is indeed the case, as Figure \ref{fig:gauss} demonstrates for $d=8$: the figure shows the direct estimates for $\ell(q)$ obtained using the RMC algorithm with $2000$ and $4000$ realisations. The differences between the results for $q \lesssim -d/2$ hints at their inaccuracy, as does an examinination of the variance of these estimates. 
More obviously, the estimates fail to capture the rapid growth of $\ell(q)$ as $ q \to -d$. Once again, we can invoke proposition \ref{prop:elltm} to remedy this problem, at least partially. Applying the RMC algorithm to the matrices $A^{-1}/|\det A|^{1/q}$ to estimate $\tilde \ell(q)$, then use the equality $\ell(q) = \tilde \ell (-q-d)$ gives a much better approximation for $\ell(q)$ in the range $-d \le q \lesssim -d/2$. Thus the approximation for $\ell(q)$ obtained in this manner with $K=2000$ and $K=4000$ are very close to one another and provide a satisfactory estimate for $q$ close to $q=-d=-8$, though the divergence at $-d$ remains difficult to capture. Note that the large-$q$ asymptotics \eqn{gaussian} has been verified to apply to the symmetric Gaussian matrices with $d=8$ considered here; it is easy to check directly from \eqn{gausRMCo} that it is satisfied for matrices with iid Gaussian  entries.

\section{Discussion} \label{conc}

Motivated by the key role played by the large-deviation statistics of Lagrangian stretching in controlling several aspects of fluid mixing, this paper examines the generalised Lyapunov exponents of products of independent random matrices. 
Such products appear in this context when the renewing flows, that is, sequences of simple iid steady flows, are used to model complex fluid motion. Products of random matrices appear of course in many other areas such as disordered media and wave localisation. 

The main aim of the paper is to present and test a reliable numerical procedure for the evaluation of the generalised Lyapunov exponents. The procedure proposed remedies the undersampling problem that affects the straightforward, brute-force Monte Carlo estimation by introducing a resampling step which ensures that the variance of the estimate scales linearly with $N$, the (large) number of matrix multiplications, rather than exponentially. The algorithm chosen, which we term Resampled Monte Carlo, is a particularly simple example of sequential importance sampling; its efficiency could be improved, e.g.\ by resampling every few iterations only, or by modifying the resampling method \citep[see][for  alternative approaches]{liu01}. 

In particular, resampling methods can be devised on the model of the PERM method used in the simulation of polymer chains \citep[][and references therein]{gras97,gras02,hsu-et-al}. In this method, the resampling is carried out only for realisations whose weight (i.e., contribution to the estimate of $\E \|X_n\|^q$ in our context) exceeds or falls below two chosen thresholds. If a weight exceeds the upper threshold, the realisation is cloned a number of times, with the weight of each clone divided accordingly; if a weight falls below the lower threshold, the realisation is either pruned with probability $1/2$ or has its weight doubled. We have implemented a method of this type, using also a random pruning to keep the number of realisations constant. The results are similar to those obtained with the RMC method, but the PERM-like method proved somewhat slower in the examples we considered. However, we have made no attempt at optimising the choice of the parameters that appear in the method (threshold values and number of clones). The PERM method has the advantage of potentially alleviating the problem of sample impoverishment which occurs for large $N$ when most of the realisations share the same early history. This problem does not appear to be serious for the computations of the generalised Lyapunov exponents of the matrix ensembles we treat in this paper, because convergence is achieved at moderately large $N$. Perhaps a more significant advantage of the PERM method in our context is that it can be implemented in a depth-first version, where the successive matrix multiplications are performed for a single realisation at a time. The drastically reduced memory requirements of depth-first approaches make them suitable for the computations of the generalised Lyapunov exponents of very large matrices. 

We have emphasised that the RMC method, and indeed all methods based on `go-with-the-winners' strategies have resampling strategies that are tailored to a particular value of $q$. When estimates of $\ell(q)$ are desired over a range of values of $q$, the computational efficiency could be improved by using the same ensemble, and hence the same resampling, for several values of $q$ within a narrow interval, rather than a separate ensemble for each value of $q$. We do not pursue these improvements here, preferring to focus on the simple version of the algorithm which can be easily analysed and already provides a dramatic improvement compared with the brute-force Monte Carlo used by many authors. 
 
In addition to providing a numerical method for the evaluation of the generalised Lyapunov exponents,  the paper dicusses some of their properties and, in particular, the relationship between the exponents associated with an ensemble of matrices $A$ and those associated with the corresponding ensemble of matrices $A^{-1}/|\det A|^{1/q}$. This relationship is useful in practice to estimate $\ell(q)$ for negative $q$, when a direct application of 
our algorithm to the matrices $A$ can be inaccurate. We also examine the asymptotic form of $\ell(q)$ for  $|q| \gg 1$ and illustrate, on a  specific example, how this form can be obtained by semi-analytical means. Asymptotic results of this type usefully complement the direct numerical estimates of $\ell(q)$ which require very large samples as $|q|$ increases.

We conclude this paper by indicating a few possible extensions of  the work reported. While the  paper focuses on the largest generalised Lyapunov exponents, which encode the large-deviation statistics of the largest finite-$N$ Lyapunov exponents, analogous statistics for lower Lyapunov exponents (proportional to the logarithm of smaller singular values of $A_N \cdots A_1$) are of interest.  It would therefore be useful to develop an  efficient numerical method to evaluate the corresponding generalised Lyapunov exponents. 
Work along these lines is currently in progress. Another useful extension concerns product of correlated matrices. The algorithm presented in section \ref{rmc} uses the independence of the matrices only to take a sequential approach, and hence it can be employed for dependent matrices provided that the dependence is on the past only, that is, that $A_n$ remain independent of $A_k$ for $k > n$.
More involved dependence would require a rethink of the algorithm. 
Since the literature on fluid mixing literature makes extensive use of white-in-time velocity fields as an alternative to renewing flows, it would also be desirable to develop methods for the efficient evaluation of generalised Lyapunov exponents in the context of linear stochastic differential equations \citep[see][for recent analytical results]{chet-et-al}. Finally, we note that the methods discussed in this paper apply to large matrices ($ d \gg 1$) and so could be employed to study the large-deviation statistics of discretised infinite-dimensional systems as arise, for instance, in the problem of passive scalar decay \citep{v06a}.


\begin{acknowledgments} 
The author acknowledges the support of a Leverhulme Research Fellowship and thanks A. Tzella for useful discussions. The anonymous referees are thanked for their useful suggestions. 
\end{acknowledgments}

\appendix

\section{Proof of proposition 2} \label{app:prop}

We first obtain the adjoint in $L_2(S^{d-1})$ of $\L_q$. Denoting by $\d \n$ the volume element on $S^{d-1}$, we consider two arbitrary functions $v(\n)$ and $w(\n)$ and compute 
\begin{eqnarray*}
\int_{S^{d-1}} w(\n) \L_q v(\n) \, \d \n &=& \E \int_{S^{d-1}} \| A \n \|^q w(\n) v(A \n/\|A \n \|) \, \d \n \\
&=&  \E \int_{S^{d-1}} \| A^{-1} \n' \|^{-q-d} w(A^{-1} \n'/\|A^{-1} \n\|) v(\n')  \, \d \n'/|\det A|,
\end{eqnarray*}
where we have changed integration variable from $\n$ to $\n'=A \n/\| A \n \|$ and used that $\d \n = \| A \n \|^d \d \n'/|\det A| = \| A^{-1} \n'\|^{-d}\d \n' / |\det A|$ \citep[cf.][]{zeld-et-al}. This gives the adjoint of $\L_q$ as
\beq \lab{Lqdag}
\L_q^\dagger = \tilde \L_{-q-d},
\eeq
where the operator $\tilde \L_q$ is defined by
\beq \lab{tildeLq}
\left(\tilde \L_q v \right)(\n) =  \E \frac{\|A^{-1} \n\|^q }{|\det A|} v \left( A^{-1} \n / \| A^{-1} \n \| \right).
\eeq
Note that $\tilde \L_q$ is equivalent to $\L_q$, with the matrices $A$ replaced by  $A^{-1}/|\det A|^{1/q}$. Thus, according to proposition 1, $\tilde \ell(q)$ defined in \eqn{tilell} is the logarithm of the largest eigenvalue of $\tilde \L_q$. The first part of proposition 2 follows from the fact that $\L_q$ and $\L_q^\dagger = \tilde \L_{-q-d}$ have the same spectrum. The second part is a particular case of the first for $\det A =1$.


We establish the third part of the proposition by showing that
\[
\ell(q)=\ell^-(q) \inter{for symplectic matrices}.
\]
Recall first that the matrices $A$ are symplectic iff
$d$ is even, and
\[
A^\mathrm{T} \mathbb{J} A = \mathbb{J}, \inter{where} \mathbb{J} = \left( 
\begin{array}{cc}
0 & \mathbb{I} \\
- \mathbb{I} & 0 \end{array}
\right),
\]
with $\mathbb{I}$ the $d/2 \times d/2$ identity matrix. Now, we define the operator $\mathcal{J}$ acting on functions on $S^{d-1}$ according to
\[
\left(\mathcal{J} v\right)(\n) = v (\mathbb{J} \n)
\]
and compute
\begin{eqnarray*}
\left(\mathcal{J} \L_q v \right) (\n) &=& \E \| A \mathbb{J} \n \|^q v \left( A \mathbb{J} \n / \| A \mathbb{J} \n \| \right) 
= \E \| \mathbb{J} A^{-\mathrm{T}} \n  \|^q v \left( \mathbb{J} A^{-\mathrm{T}} \n / \| \mathbb{J} A^{-\mathrm{T}} \n \| \right) \\
&=& \E \| A^{-\mathrm{T}} \n  \|^q v \left( \mathbb{J} A^{-\mathrm{T}} \n / \| A^{-\mathrm{T}} \n \| \right) 
= \E \| A^{-\mathrm{T}} \n \|^q  \left( \mathcal{J} v \right) \left( A^{-\mathrm{T}} \n / \| A^{-\mathrm{T}} \n \| \right) \\
&=& \left(\L^\mathrm{-T}_q \mathcal{J} v \right) (\n),
\end{eqnarray*}
where $\L^{-\mathrm{T}}_q$ is the analogue of $\L_q$ with $A^{-\mathrm{T}}$ replacing $A$. This computation shows that $\L_q$ and $\L^{-\mathrm{T}}_q$ have the same spectrum, hence $\ell(q)=\ell^{-\mathrm{T}}(q)$. The results follows from observing that $\ell^{-\mathrm{T}}(q)=\ell^-(q)$ since they can be expressed as expectation of the largest singular values of $A^{-1}_N\cdots A^{-1}_1$ and $A^{-\mathrm{T}}_N\cdots A^{-\mathrm{T}}_1$, which coincide.

\section{Pseudocode for the RMC method} \label{pseudo}

We give below a pseudocode for the RMC algorithm. The notation is as in section \ref{algo} except for the omission of the superscripts $(k)$ and subscripts $n$ when these are unnecessary for the numerical implementation. The variables $\gamma^{(k)}$ are introduced to perform the random resampling \eqn{algo2}--\eqn{algo3} using the uniformly distributed random variables $\epsilon$.
\bigskip

\begin{tabular}{ll}
fix $q$ & \\
$E^{(k)}=(1,0,\cdots,0)^\mathrm{T}, \, k=1,\cdots,K$ & (unit vectors in $\mathbb{R}^d$) \\
\textbf{for} $n=1$ \textbf{to} $N$ & (loop over iterations)\\
\mbox{~~~~} \textbf{for} $k=1$ \textbf{to} $K$ & (loop over realisations)\\
\mbox{~~~~} \mbox{~~~~} draw random matrix $A$ \\
\mbox{~~~~} \mbox{~~~~} $\hat E^{(k)}=A E^{(k)}/\|A E^{(k)}\|$ & \\
\mbox{~~~~} \mbox{~~~~} $\alpha^{(k)}=\|A E^{(k)}\|^q$ & \\
\mbox{~~~~} \textbf{end} & \\
\mbox{~~~~}$\gamma^{(k)}=\sum_{l=1}^k \alpha^{(l)}, \, k=1,\cdots,K$ & \\
\mbox{~~~~}$\beta_n = \gamma^{(K)}$ & \\
\mbox{~~~~} \textbf{for} $k=1$ \textbf{to} $K$ & (resampling)\\
\mbox{~~~~} \mbox{~~~~} draw $\epsilon$ uniformly in $[0,\beta_n]$ & \\
\mbox{~~~~} \mbox{~~~~} $j=\min\{l \in \{1,\cdots, K\}\, |\, \gamma^{(l)}-\epsilon \ge 0\}$  & \\
\mbox{~~~~} \mbox{~~~~} $E^{(k)}=\hat E^{(j)}$ & \\
\mbox{~~~~} \textbf{end} & \\
\textbf{end} & \\
$\ell = \left(\sum_{n=1}^N \log \beta_n \right)/N - \log K$ & (estimate of $\ell(q))$ \\
\textbf{end} & 
\end{tabular}

\section{Bounds on $\kappa$} \label{bound}

Starting from \eqn{eigw}, we write $\kappa$ as
\beq \lab{kapap}
\kappa = \sup_A \left( \log \|A \n \| + w(A \n/\|A \n\|) \right) - w(\n),
\eeq
which holds for any $\n$. Taking $\n = \n_w$, where $\n_w$ maximises $w$ gives
\[
\kappa \le \sup_A \log \| A \n_w \|,
\]
and in particular
\[
\kappa \le \sup_{\n,\, A} \log \| A \n \| = \log \sup_{\n,\, A} \| A \n \| = \log \sup_A \| A \|.
\]
This is also obvious from the fact that $\lambda_1 \le \sup_{\n, \, A} \|A \n \|^q$. On the other hand, denoting by $\n_*$ and $A_*$ the maximisers of $\| A \n \|$, and evaluating \eqn{kapap} at $\n = \n_*$, we have that
\[
\kappa \ge \log \| A_* \n_*\| + w(A_* \n_*/\|A_* \n_*\|) - w(\n_*) = \log \sup_A \|A\| +  w(A_* \n_*/\|A_* \n_*\|) - w(\n_*).
\]
A consequence is that $w(A_* \n_*/\|A_* \n_*\|) \le w(\n_*)$. In the special case where 
\[
\frac{A_* \n_*}{\|A_* \n_*\|} = \n_*,
\]
the two inequalities obtained imply that
\[
\kappa = \log \sup_A \| A \|.
\]









\bibliographystyle{apsrev4-1long}
\bibliography{mybib}

\end{document}